\documentclass[aps,pra,superscriptaddress,showpacs,twocolumn]{revtex4-2}
\usepackage{bm}
\usepackage{amsmath}
\usepackage{amssymb}
\usepackage{slashed}
\usepackage{float}
\allowdisplaybreaks
\usepackage{subcaption}
\usepackage{dcolumn}
\usepackage{graphicx}
\newcolumntype{.}{D{x}{}{-1}}

\newcolumntype{w}[1]{D{.}{.}{#1}}
\newcolumntype{L}{>{$}l<{$}}
\newcommand{\Za}{Z\alpha}
\usepackage{nicefrac}

\newcommand{\lbr}{\langle}
\newcommand{\rbr}{\rangle}

\begin{document}

\title{Hyperfine splitting in $\bm{^{6,7}}$Li$\bm{^+$}}

\author{Krzysztof Pachucki}
\affiliation{Faculty of Physics, University of Warsaw,
             Pasteura 5, 02-093 Warsaw, Poland}

\author{Vojt\v{e}ch Patk\'o\v{s}}
\affiliation{Faculty of Mathematics and Physics, Charles University,  Ke Karlovu 3, 121 16 Prague
2, Czech Republic}

\author{Vladimir A. Yerokhin}
\affiliation{Max–Planck–Institut f\"ur Kernphysik, Saupfercheckweg 1, 69117 Heidelberg, Germany}

\begin{abstract}

We present a detailed derivation of the QED effects of order $\alpha^7\,m$ to
the hyperfine structure (hfs) of the $^3S$ states of
heliumlike ions and perform numerical calculations for $^6$Li$^+$ and $^7$Li$^+$.
By comparing the theoretical point-nucleus results with the measured hfs of Li$^+$,
we determine the nuclear-structure contribution parametrized in terms of
the effective Zemach radius.
Using the experimental hfs results for Li$^+$,
we obtain accurate predictions for the hfs of $^6$Li$^{2+}$
and $^7$Li$^{2+}$, for which no experimental data is available so far.
By examining the normalized differences of the hfs of Li$^{+}$ and Li and
of the corresponding isotope-shift differences, we test the consistency
of the hfs measurements in $^{6,7}$Li$^{+}$ and $^{6,7}$Li.

\end{abstract}

\maketitle

\section{Introduction}

The hyperfine structure (hfs) of atomic levels with vanishing orbital angular momentum
arises from the interaction between the nuclear spin
and the intrinsic angular momentum (spin) of the electrons within an atom.
The energy level of an atomic $S$ state can be conveniently represented
as a sum of the centroid energy level  $E_{\rm cent}$, the
magnetic dipole hyperfine structure $E_{M1}$
and the electric quadrupole hyperfine structure $E_{E2}$,
\begin{align}
E =&\, E_{\rm cent} + E_{M1} + E_{E2}
 \nonumber \\ \equiv &\,
E_{\rm cent} + A\,
 \lbr \vec I\cdot\vec S\rbr + B\, \lbr(I^i I^j)^{(2)}(S^i S^j)^{(2)}\rbr\,,
\end{align}
where $A$ and $B$ are the so-called hyperfine constants,
$\vec S$ is the total spin of electrons, $\vec I$ is the nuclear spin operator, $(a^ia^j)^{(2)}$ denotes the
irreducible second-rank tensor, and the summation over repeated indices is implicit.
The matrix elements are given by
\begin{align}
\lbr \vec{I}\cdot \vec{S}\rbr = \big[ F(F+1) - I(I+1) -S(S+1)\big]/2\,,
\end{align}
and
\begin{align}
 \lbr(I^i I^j)^{(2)}(S^i S^j)^{(2)}\rbr  = &\,
  \lbr\vec{I}\cdot\vec{S}\rbr^2 + \lbr\vec{I}\cdot\vec{S}\rbr/2
\nonumber \\ &
  -
I(I+1)S(S+1)/3\,.
\end{align}
We note that the electric quadrupole structure
is present only for the electron states with the total momentum
$J=S > 1/2$; in particular, it vanishes for the ground state of atomic Li and
Li$^{2+}$.

In the present work we are interested in the magnetic dipole hfs.
In order to obtain
experimental results for the $A$ constant from the measured spectra, one needs
to eliminate the quadrupole structure by combining
several hfs transitions.
Using the most accurate experimental results summarized in Table~\ref{tab:exper},
we get for the Li$^+$ ion
\begin{align}\label{eq:aexp:0}
A_\mathrm{exp}({}^6\mathrm{Li^+}) = &\ \frac16\,\nu_{0-1} + \frac{5}{12}\,\nu_{1-2}\,,\\
A_\mathrm{exp}({}^7\mathrm{Li^+}) = &\  \frac16\,\nu_{1/2-3/2} + \frac{3}{10}\,\nu_{3/2-5/2}\,,
\label{eq:aexp:1}
\end{align}
where $\nu_{F-F'}$ are the measured $F-F'$ transition energies.
For atomic Li, the quadrupole structure is absent, so we have just
\begin{align}
A_\mathrm{exp}({}^6\mathrm{Li}) = &\ \frac23\,\nu_{1/2-3/2} \,,\\
A_\mathrm{exp}({}^7\mathrm{Li}) = &\  \frac12\,\nu_{1-2}\,.
\end{align}

\begin{table}
\caption{Experimental hfs intervals in $\mathrm{Li^+}$ and $\mathrm{Li}$, in MHz.
\label{tab:exper}}
\begin{ruledtabular}
\begin{tabular}{llcc}
\multicolumn{1}{l}{System} &
\multicolumn{1}{l}{Interval} &
        \multicolumn{1}{c}{Experiment}
        &
        \multicolumn{1}{c}{Ref.}
\\
\hline\\[-5pt]
${}^6\mathrm{Li}^+$ & $2^3 S_1^{0-1}$ & 3001.783(12) & \cite{sun:23}\\
${}^6\mathrm{Li}^+$ & $2^3 S_1^{1-2}$ & 6003.619(11) & \cite{sun:23}\\
${}^7\mathrm{Li}^+$ & $2^3 S_1^{1/2-3/2}$ & 11\,890.088(65) & \cite{guan:20}\\
${}^7\mathrm{Li}^+$ & $2^3 S_1^{3/2-5/2}$ & 19\,817.696(42) & \cite{guan:20}\\
${}^6\mathrm{Li}$   & $2^2 S_{1/2}^{1/2-3/2}$ & 228.205\,259\,0(30) & \cite{beckmann:74}\\
${}^7\mathrm{Li}$   & $2^2 S_{1/2}^{1-2}$ & 803.504\,086\,6(10) & \cite{beckmann:74}\\
\end{tabular}
\end{ruledtabular}
\end{table}

%

\section{Hfs theory of light hydrogen-like ions}
\label{sec:hydr}

We start with summarizing the existing theory for the magnetic dipole hfs of $S$ states
of hydrogen-like atoms. To the leading order in the fine-structure constant $\alpha$, it is given by
\begin{align}\label{eq:01}
E_F =
\frac{4}{3}(Z\alpha)^4\frac{m_\mathrm{r}^3}{m\,M}\,g \, \lbr \vec{I}\cdot \vec{S}\rbr\,,
\end{align}
where $m_\mathrm{r} = mM/(m+M)$ is the reduced mass,
$m$ and $M$ are the mass of the electron and nucleus,
respectively,
$Z$ is the nuclear charge number, and the nuclear $g$-factor is defined as
\begin{align}
\vec\mu = \frac{Z\,e}{2\,M}\,g\,\vec I\,,
\end{align}
where $\vec\mu$ is the nuclear magnetic moment operator.
The leading-order hfs contribution in Eq.~(\ref{eq:01}) is also known as the Fermi energy.

The complete hyperfine structure of $S$ states in hydrogenic systems
is represented as an expansion in terms of $\alpha$,
\begin{align}\label{eq:13}
E_{M1} =&\ E_F\,(1 + \delta)\,,
\end{align}
where
\begin{align} \label{07}
\delta  =&\
\kappa+ \delta^{(2)}+ \delta^{(3)} + \delta^{(4)}
 + \delta^{(2)}_\mathrm{rec.rel}
 + \delta_\mathrm{struct}\,.
\end{align}
Here, $\kappa $ is the magnetic moment anomaly of a free electron, $\kappa = \alpha/(2\pi) + O(\alpha^2)$,
$\delta^{(i)}$ are QED corrections of order $\alpha^i E_F$,
$\delta^{(2)}_\mathrm{rec.rel}$ is
the relativistic recoil correction of order $\alpha^2\,E_F$,
and $\delta_\mathrm{struct}$ is the nuclear structure correction.

The results for the QED corrections for an $S$ state and a point and infinitely heavy nucleus
are given by~\cite{eides:01,tiesinga:21:codata18}
\begin{align}\label{08}
\delta^{(2)}&\  =\frac{3}{2}\,(Z\,\alpha)^2 + \alpha\,(Z\,\alpha) \Bigl(\ln 2-\frac{5}{2}\Bigr)\,,\\
\delta^{(3)}&\  = \frac{\alpha\,(Z\,\alpha)^2}{\pi}\!
\bigg[\!-\frac{8}{3}\ln(Z\,\alpha)\Bigl(\ln(Z\,\alpha)-\ln 4+\frac{281}{480}\Bigr)
\nonumber \\ &\
+ 17.122\,338\,751\,3-\frac{8}{15}\,\ln 2+\frac{34}{225}\biggr]
\nonumber \\ &\
+ \frac{\alpha^2\,(Z\,\alpha)}{\pi}\,0.770\,99(2) \,, \label{09}\\
\delta^{(4)} &\  =\frac{17}{8}\,(Z\,\alpha)^4 +\alpha\,(Z\,\alpha)^3\,\bigg[\Big(\frac{547}{48}-5\,\ln 2 \Big)\,\ln(Z\,\alpha)
\nonumber \\ &\
+G_{\rm SE}^{(4)}(\Za) +\frac{13}{24}\,\ln 2 + \frac{539}{288} \biggr]
\nonumber
\\ &\
-\frac{\alpha^2\,(Z\,\alpha)^2}{\pi^2}\,\biggl[\frac{4}{3}\,\ln^2(Z\,\alpha)+1.278\,\ln(Z\,\alpha) +10.0\pm2.5\biggr]
  \nonumber \\ & \
\pm  \frac{\alpha^3\,(Z\,\alpha)}{\pi^2}\,.
\label{10}
\end{align}
Here, $G_{\rm SE}^{(4)}(\Za)$ is the one-loop self-energy correction which needs to be calculated numerically.
For Li, we use the result from Ref.~\cite{yerokhin:08:prl} of $G_{\rm SE}^{(4+)}(3\alpha) = -4.587\,5(1)$
which includes  higher orders in $Z\,\alpha$ for $Z=3$. Furthermore,
the last term in $\delta^{(4)}$ represents the estimate of the
unknown three-loop binding QED correction.

The relativistic recoil correction was derived in Ref.~\cite{Bodwin:88}.
It has a finite point-nucleus limit and is given by
\begin{align}
\delta^{(2)}_\mathrm{rec.rel} =&\ (Z\,\alpha)^2\,\frac{m_\mathrm{r}^2}{m\,M}\,\biggl\{-\bigg[-6 + \frac{7}{2}g + \frac{14}{g}\biggr]\,\frac{\ln(Z\,\alpha)}{4}
\nonumber \\ &\ \hspace{-10ex}
-\biggl[-2 + \frac{11}{2}g + \frac{46}{g}\biggr]\,\frac{\ln2}{4} + \frac{1}{36}\biggl[ -51 + \frac{31}{2}g + \frac{300}{g}\biggr]\biggr\}\,.
\end{align}

The last term in Eq.~(\ref{07}) is the nuclear-structure contribution $\delta_{\rm struct}$.
Its dominant part is the elastic contribution of order $\alpha\, E_F$, which is parametrized in terms of
the Zemach radius $r_Z$,
\begin{align}
\delta^{(1)}_\mathrm{struct}(\mbox{\rm elastic}) =&\ -2\,Z\,\alpha\,m_\mathrm{r}\,r_Z\,,
\end{align}
where
\begin{equation}
r_Z = \int d^3 r_1 \int d^3 r_2 \,\rho_E(\vec r_1)\,\rho_M(\vec r_2) \,|\vec r_1 - \vec r_2|\,,
\end{equation}
and $\rho_E$ and $\rho_M$ are the Fourier transforms of the electric and magnetic form factors
of the nucleus, normalized to unity.

There are many further contributions to $\delta_{\rm struct}$, both of the elastic and the inelastic kind.
So far there is no established theory for calculating the inelastic nuclear effects in hfs for
a compound nucleus. For this reason, we parametrize the {\em whole} nuclear structure contribution
in terms of the {\em effective} Zemach radius $\tilde{r}_Z$, which is, by definition,
\begin{align}\label{eq:rZtilde}
\delta_\mathrm{struct} \equiv -2\,Z\,\alpha\,m_\mathrm{r}\,\tilde{r}_Z\,.
\end{align}
It should be noted that the definition of the nuclear
structure contribution (and, therefore, 
the effective Zemach radius) is not unique. In particular, in our previous study of He$^+$
\cite{patkos:23} we separated out from $\delta_\mathrm{struct}$ the nuclear recoil correction
$\delta^{(1)}_\mathrm{rec}$ and the elastic higher-order nuclear
contribution $\delta^{(2)}_\mathrm{nuc}$. Moreover, many previous studies (among them,
Refs.~\cite{qi:20,sun:23}) disregarded all higher-order nuclear contributions, thus making 
no difference between the elastic Zemach radius $r_Z$ and the effective Zemach radius $\tilde{r}_Z$.
In the present work, we define the nuclear structure contribution by Eq.~(\ref{07}),
where we separated out only those corrections that have a well-defined point-nucleus
limit.

%

\section{Hfs theory of light few-electron atoms}

For the $S$ states, the leading-order magnetic dipole hyperfine structure is
given by
\begin{align}\label{eq:EF}
E_F \equiv \lbr V_F \rbr    =&\  \frac{4\,\pi\,Z\alpha}{3\,m\,M}\,g\,
\Big< \vec I\cdot \sum_a  \vec s_a\,\delta^3(r_a)\Big>\,.
\end{align}
The matrix element in the above expression is assumed to be calculated for a finite nuclear mass
and thus implicitly contains the reduced mass prefactor.
For a hydrogen-like atom, Eq.~(\ref{eq:EF}) reduces to Eq.~(\ref{eq:01}).
It is often convenient to separate out from $E_F$ the dependence on the total angular momentum by introducing
the hyperfine constant $A_F$ which does not depend on the hyperfine state,
\begin{align}
E_F = A_F\,\langle \vec I\cdot\vec S\rangle \,.
\end{align}
The complete magnetic dipole hyperfine structure is expressed analogously to Eq.~(\ref{eq:13})
\begin{align}
E_{M1} =  E_F (1+\delta)\,,
\end{align}
where
\begin{align} \label{eq:delta}
\delta =   \kappa& + \delta^{(2)}\! + \delta^{(3)} + \delta^{(4)}
 + \delta^{(2)}_\mathrm{rec}
 + \delta_\mathrm{struct}\,.
\end{align}

Calculations of the leading-order
magnetic dipole hfs in helium and heliumlike atoms are presently well established
\cite{morton:06:cjp}. The leading QED correction of order $\alpha^2E_F$, $\delta^{(2)}$,
was derived and calculated for helium in our previous investigations
\cite{pachucki:01:jpb,pachucki:12:hehfs}.
For the Li$^+$ and Be$^{2+}$ ions, similar calculations were carried out in Refs.~\cite{qi:20,sun:23,qi:23}.
The higher-order QED correction $\delta^{(3)}$ was reported by us for helium atom in
Ref.~\cite{patkos:unpunlished}. In the next section, we
present the detailed derivation of formulas and perform numerical calculations
of $\delta^{(3)}$ for Li$^+$. The higher-order QED correction
$\delta^{(4)}$ is approximated by the hydrogenlike value in Eq.~(\ref{10}).
While it is a relatively small correction, its uncertainty
will define the overall uncertainty of our theoretical predictions for the point-nucleus hfs.

The complete relativistic recoil correction $\delta^{(2)}_\mathrm{rec}$  for few-electron systems is presently unknown.
In this work we approximate it by a sum of the relativistic recoil correction
for the corresponding hydrogenic ion and the mixing contribution
$\delta^{(2)}_\mathrm{rec.mix}$,
\begin{align}
\delta^{(2)}_\mathrm{rec}  = \delta^{(2)}_\mathrm{rec.rel}(\mathrm{Li}^{2+} )+ \delta^{(2)}_\mathrm{rec.mix}\,.
\end{align}
The mixing correction $\delta^{(2)}_\mathrm{rec.mix}$
is a second-order (in the magnetic moment) contribution due to mixing with closely lying excited states.
This correction is specific for the heliumlike ions, where the reference $2^3S$ state
and the first excited
$2^1S$ state are separated by
a small energy interval. Namely, for the $2^3S$ reference state, the
following mixing contribution is present:
\begin{align}
&\frac{\bigl\langle 2^3S| V_F | 2^1S\bigr\rangle\,\bigl\langle 2^1S| V_F | 2^3S\bigr\rangle}{E(2^3S)-E(2^1S)}  \nonumber \\ &
= I^i\,I^j \,
\frac{\bigl\langle 2^3S| V_F^i | 2^1S\bigr\rangle\,\bigl\langle 2^1S| V_F^j | 2^3S\bigr\rangle}{E(2^3S)-E(2^1S)}   \nonumber \\ &
\sim
\frac{i}{2}\,\epsilon^{ijk}\,I^k
\,\frac{\bigl\langle 2^3S| V_F^i | 2^1S\bigr\rangle\,\bigl\langle 2^1S| V_F^j | 2^3S\bigr\rangle}{E(2^3S)-E(2^1S)}  \nonumber \\ &
=A_F\,\vec I\cdot\vec S\, \delta^{(2)}_\mathrm{rec.mix}\,,
\end{align}
where we retained only the contribution to the magnetic dipole hfs.
Despite being second order in the electron-nucleus mass ratio, this correction is significant because
of the small energy difference in the denominator.
We note that a similar correction contributes to the electric quadrupole structure
and is present even when the quadrupole moment of the nucleus is zero.

The nuclear-structure contribution $\delta_\mathrm{struct}$ cannot be calculated at present.
Instead, we will extract it from the experimental value of hfs in Li$^+$.
It is important that, with high accuracy, the nuclear structure contribution expressed in
terms of $\delta_\mathrm{struct}$ is the same for Li$^{2+}$, Li$^{+}$, and Li.
This fact will allow us to predict the hyperfine structure of Li$^{2+}$ and to cross-check
the experimental results for Li$^+$ against those for Li.

%
\section{QED correction of order $\bm{\alpha^3E_F}$}

We now turn to the derivation of
the QED contribution of the order of $m\alpha^7$ ($\equiv \alpha^3E_F$) to the
magnetic dipole hyperfine
structure of triplet $S$ states in few-electron atoms. The $m\alpha^7$ contribution to $E_{M1}$ consists of
the photon-exchange terms (no radiative loops), the one-loop self-energy, the one-loop vacuum polarization, and the two-loop QED effects.
It  can be represented in terms of the first- and second-order matrix elements as
\begin{align}\label{1}
&\ E^{(7)}
\equiv E_F\,\delta^{(3)}
=  E_L + 2\,\bigg\langle H^{(4)}_{\rm hfs}\frac{1}{(E_0-H_0)'}\,H^{(5)}\bigg\rangle\nonumber\\
&\ +2\,\bigg\langle H^{(5)}_{\rm hfs}\frac{1}{(E_0-H_0)'}\,H^{(4)}\bigg\rangle + \langle H^{(7)}_{\rm hfs}\rangle
+E_\mathrm{2loop}\,,
\end{align}
where $E_L$ is the low-energy Bethe-logarithm-type contribution;
$H^{(4)}_{\rm hfs}$,  $H^{(5)}_{\rm hfs}$, and  $H^{(7)}_{\rm hfs}$
are the effective hfs Hamiltonians of order $m\alpha^4$, $m\alpha^5$, and $m\alpha^7$, respectively;
$H^{(4)}$ is the Breit Hamiltonian,
$H^{(5)}$ is the effective QED Hamiltonian of order $m\alpha^5$,
and $H_0$ and $E_0$ are the nonrelativistic Hamiltonian and its reference-state eigenvalue. The lowest-order hfs Hamiltonian
$H^{(4)}_{\rm hfs}$ is given by Eqs.~(5)-(11) of Ref.~\cite{pachucki:12:hehfs}, where one should put the electron
anomalous magnetic moment (amm) to zero. The next-order hfs Hamiltonian $H^{(5)}_{\rm hfs}$ is the
amm correction to $H^{(4)}_{\rm hfs}$ and is immediately obtained from
Eqs.~(5)-(11) of Ref.~\cite{pachucki:12:hehfs}.
The Breit-Pauli Hamiltonian is well known and given, e.g., by Eq.~(7) of Ref.~\cite{yerokhin:10:helike}.
The effective QED Hamiltonian of order $m\alpha^5$ is
\begin{align}
H^{(5)} = &\ \bigg(\frac{5}{6} -\frac{1}{5} + \ln\frac{\alpha^{-2}}{2\lambda}\bigg)\,\frac{4\,\alpha^2 Z}{3\,m^2}\,\big[\delta^3(r_1)+\delta^3(r_2)\big]
\nonumber \\ &\
-\frac{7\,\alpha^2}{3\,\pi\,m^2}\,\frac{1}{r^3}
+ H^{(5)}_{\rm fs}\,, \label{H5}
\end{align}
where $r = |\vec{r}_1-\vec{r}_2|$, $\lambda$ is the low-energy cutoff of photon momenta, and
$H^{(5)}_{\rm fs}$ is the amm correction to the spin-dependent Breit-Pauli Hamiltonian,
given by Eq.~(14) of Ref.~\cite{yerokhin:10:helike}.
$E_\mathrm{2loop}$ is the two-loop contribution which has
the same form as in hydrogen-like atoms, see the last term in Eq. (\ref{09}),
and is given by
\begin{align}\label{2loop}
 E_{\rm 2loop} = E_F\,\frac{\alpha^2\,(Z\,\alpha)}{\pi}\,0.770\,99(2)\,.
\end{align}

The derivation of the low-energy contribution $E_L$ and the first-order $m\alpha^7$ effective Hamiltonian
$H^{(7)}_{\rm hfs}$ is presented below.

\subsection{Low-energy contribution $\bm{E_L}$}\label{lowen}

In this section, it will be convenient to use the atomic units,
and pull out the overall prefactor $\alpha^7$. Also, in the rest of the paper we will
set the electron mass to unity, $m=1$, unless specified otherwise.

In the low-energy region the momentum of the virtual photon $k$ is of the order $k\approx\alpha^2$.
The corresponding low-energy hfs contribution comes from the perturbation of the Bethe logarithm by
the Fermi contact interaction operator
\begin{equation}\label{eq:VF}
{\cal V}_F = \frac{g}{M} \frac{2\,\pi\,Z}{3}\,\vec I\cdot\vec S\,\big[\delta^3(r_1) + \delta^3(r_2)\big]\,.
\end{equation}

The low-energy contribution $E_L$ is
\begin{align}
E_{L} &\ = \frac{2}{3\,\pi} \int_0^{\lambda} dk\,k P_{L}(k)\,,
\end{align}
where
\begin{align}\label{eq:PL}
&\ P_{L}(k)  = 2 \left< {\cal V}_F \frac1{(E_0-H_0)'}\, \vec{P}\,  \frac1{E_0-H_0-k} \, \vec{P} \right>
 \\
&\ +  \left< \vec{P}\,  \frac1{E_0-H_0-k} \,\bigl[ {\cal V}_F - \lbr {\cal V}_F\rbr\bigr]  \frac1{E_0-H_0-k} \,\vec{P} \right>\,,
\nonumber
\end{align}
and $\vec{P} = \vec{p}_1+ \vec{p}_2$ is the total momentum operator.
The large-$k$ expansion of $P_{L}(k)$ is
\begin{align}\label{eq:as}
k P_{L}(k) &\ = A + \frac{B}{\sqrt{k}} + \frac{C\,\ln k}{k} + \frac{D}{k} + \ldots\,,
\end{align}
where $A$, $B$, $C$, and $D$ are the asymptotic constants summarized in Appendix~\ref{app:asympt}.
The $\lambda$-dependent part of $E_L$ is separated as
\begin{equation}\label{ELl}
E_L =  E_L' + \frac{2}{3\,\pi}\bigg(\frac{C}{2}\ln^2\lambda + D\,\ln\lambda\bigg)\,,
\end{equation}
where the finite $\lambda$-independent part is expressed as
\begin{align}
E'_{L}
= &\ \frac{2}{3\,\pi} \biggl\{ \int_0^K kdk\,P_{L}(k) + \int_K^{\infty} dk\,
\biggl[ k P_{L}(k)
 \nonumber \\ &
- A  - \frac{B}{\sqrt{k}}  - \frac{C\ln k}{k}  - \frac{D}{k}\biggr]
 \nonumber \\ &
 - \biggl[ A\,K + 2\,B\,\sqrt{K} + \frac{C}{2}\ln^2K + D\,\ln K\biggr]\biggr\}\,, \label{EL}
\end{align}
where $K\geq1$ is a free parameter.

In order to remove the dominant $Z$ and state dependence from $E'_{L}$, it is convenient
to define the $\delta$-function-perturbed Bethe logarithm $\beta_{\delta}$ as
\begin{align}
E'_L = \beta_{\delta} \frac{Z^2}{4\pi} \big< {\cal V}_F\big>
- \frac{2}{3\pi} \Big( \frac{C}{2} \ln^2 Z^2 + D \ln Z^2 \Big)\,.
\end{align}
Defined in such a way, $\beta_{\delta}$ depends very weakly on $Z$ and its
numerical values for few-electron atoms are very close to the hydrogenic values.

\subsection{$\bm{m\alpha^7}$ Hamiltonian $\bm{H^{(7)}_\mathrm{hfs}}$}\label{H7hfs}

The effective $m\alpha^7$ Hamiltonian $H^{(7)}_{\rm hfs}$ can be represented as
\begin{align}\label{H7se}
H^{(7)}_{\rm hfs} = &\ H^{(7)}_{{\rm hfs},A} + H^{(7)}_{{\rm hfs},B}
  + \ldots \,,
\end{align}
where $\ldots$ denotes terms that are proportional to the electron-nucleus Dirac $\delta$ function, $\delta^3(r_a)$.
At the first stage of the derivation we will routinely drop such terms; the corresponding
contribution will be restored later by matching the high-$Z$ limit of the
obtained formulas to the known hydrogenic result; see Sec.~\ref{sec:matching}.
More specifically, we will omit terms proportional to $Z^3\,\delta^3(r_a)$; all other
terms proportional to $\delta^3(r_a)$ will be preserved throughout the derivation.

The first part of $H^{(7)}_{\rm hfs}$
comes from the spin-dependent terms in the generalized Breit-Pauli Hamiltonian $H_\mathrm{BP}$
that are proportional to the electron amm $\kappa$. Specifically,
\begin{align}\label{HBP}
\delta H_{\rm BP} = &\ \sum_a \kappa\,\biggl\{  \frac{Z\alpha}{2}\, \vec\sigma_a\cdot\frac{\vec r_a}{r_a^3}\times\vec\pi_a
+  \frac{e}{8}\, \{\vec{\pi}_a\cdot\vec{B}_a\,,\, \vec{\pi}_a\cdot\vec{\sigma}_a \}
\nonumber \\ &\
+ \frac{e}{16} \,[\pi^i_a\,,\,[\pi^i_a\,,\,\vec{\sigma}_a\cdot \vec{B}_a]]\biggr\}
\nonumber \\ &\
+ \sum_{a\neq b}
\frac{\alpha\,\kappa}{2\, r_{ab}^3}\vec\sigma_a\cdot\vec r_{ab}\times (\vec\pi_b -\vec\pi_a)\,,
\end{align}
where $\vec\pi = \vec p -e\,\vec A$ and
\begin{align}\label{A}
e\,\vec A(\vec r) = &\ \frac{e}{4\pi} \vec{\mu}\times\frac{\vec r}{r^3} = -Z\alpha\, \frac{g}{2\,M} \vec I\times \frac{\vec r}{r^3}\,,\\
e\,B^i(\vec r) = &\ (\nabla\times\vec A)^i = -Z \alpha\, \frac{g}{2\,M}\frac{8\pi}{3}\delta^3(r)\,I^i
\nonumber \\ &\
 + Z\alpha \,\frac{g}{2\,M}\frac{1}{r^3} \bigg(\delta^{ij} - 3\frac{r^i r^j}{r^2}\bigg) I^j\,.\label{B}
\end{align}
Performing calculations as described in Appendix \ref{app:H7} we obtain
\begin{align}\label{H7a}
H^{(7)}_{ {\rm hfs},A}
= &\ \frac{g\,Z\alpha\,\kappa}{4\,M} \,\vec I\cdot\vec S\,\bigg\{
\frac{2Z\alpha}{3\,r_1^4 }
-\frac{1}{6}\,\bigg(\frac{8\pi}{3}\,p_1^i\,\delta^3(r_1)\,p_1^i \nonumber\\
&\
-p_1^i\,\frac{1}{r_1^5}\big(r_1^2\,\delta^{ij}-3\,r_1^i r_1^j\big)\,p_1^j\bigg)
+\frac{\pi}{3}\,\Delta\,\delta^3(r_1)
\nonumber\\&\
-  \frac43 \alpha\,\frac{\vec r\cdot\vec r_1}{r^3\,r_1^3}\bigg\}
 + (1\leftrightarrow2)\,.
\end{align}
Some operators in the above expression are singular at the origin and thus are not well defined, but this ambiguity
will be eliminated by matching with the known hydrogenic result.

The second part of $H^{(7)}_{\rm hfs}$ is a middle-energy contribution 
that can be expressed in terms of slopes of form factors
and the one-loop vacuum polarization. The derivation described in Appendix \ref{app:H7} yields
\begin{align}\label{H7b}
 H^{(7)}_{ {\rm hfs},B}
 = &\ \frac{2\pi}{3}\,\frac{g\,Z\alpha}{M}\,\vec I\cdot\vec S\, \bigg[F'_1(0)+F'_2(0)
 -\frac{\alpha}{15\,\pi}\bigg]
  \nonumber\\
&\
\times
\Delta\,\delta^3(r_1)
+ (1\leftrightarrow2)\,,
\end{align}
where the slopes of form factors are given by
\begin{equation}
F'_1(0)+F'_2(0) = \frac{\alpha}{\pi}\bigg[\frac{17}{72} + \frac13\ln\frac{\alpha^{-2}}{2\lambda}\bigg]\,.
\end{equation}

\begin{table}
\caption{Expectation values of operators $Q_i$ for the $2^3S$ state of Li$^+$, in atomic units.
Singular operators $Q_{56}$ and $Q_{57}$ are defined according to Ref.~\cite{patkos:21:helamb}.
\label{tab:oprsQ1}}
\label{oprsQ1}
\begin{ruledtabular}
\begin{tabular}{llw{3.6}}
%
$Q_1 $ & $ 4 \pi \delta^3 (r_1)$   				     &          57.350354\\
$Q_3 $ & $4 \pi \delta^3(r_1)/r_2$                  	     &  27.981057     \\	
$Q_4 $ & $4 \pi \delta^3(r_1)\, p_2^2$ 	                     &  22.668097     \\
$Q_9 $ & $1/r^3$                    	                     &   0.195563  \\
$Q_{11}$ & $1/r_1^2$                	                     &   9.601760    \\
$Q_{12}$ & $1/(r_1 r_2)$            	                     &   1.472668    \\
$Q_{13}$ & $1/(r_1 r)$              	                     &   0.860969     \\
$Q_{14}$ & $1/(r_1 r_2 r)$          	                     &   0.837624   \\
$Q_{15}$ & $1/(r_1^2 r_2)$					     &               5.002281 \\
$Q_{16}$ & $1/(r_1^2 r)$					     &               4.660766           \\
$Q_{17}$ & $1/(r_1 r^2)$   					     &               0.514395            \\
$Q_{18}$ & $(\vec{r}_1\cdot\vec r)/(r_1^3 r^3)$                  &0.083179      \\
$Q_{24}$ & $p_1^i\,(r^i r^j+\delta^{ij} r^2)/(r_1 r^3)\, p_2^j$  &0.019568       \\
$Q_{28}$ & $p_1^2\,/r_1\, p_2^2$				     &           17.346919    \\
$Q_{51} $ & $4 \pi\,\vec p_1\,\delta^3(r_1)\,\vec p_1 $	         &0.051166      \\
$Q_{53} $ & $1/r_1$						                         &1.780585   \\
$Q_{56}$ & $1/r_1^3$   					                         &-102.905512  \\
$Q_{57}$ & $1/r_1^4$   					                         &271.277651  \\
$Q_{59}$ & $1/(r_1 r^3)$				                         &0.405548   \\
\end{tabular}
\end{ruledtabular}
\end{table}

\begin{table*}
\caption{Second-order matrix elements for the $2^3S$ state of
Li$^+$, in atomic units.
``Symmetry'' denotes the symmetry of the intermediate states.
\label{table:sec}
}
\begin{ruledtabular}
\begin{tabular}{llcw{5.8}}
\multicolumn{1}{l}{} &
    \multicolumn{1}{l}{} &
    \multicolumn{1}{l}{Symmetry} &
            \multicolumn{1}{c}{Value}
\\
\hline\\[-5pt]
$S_1 $&$ \Big<V_R\,\frac{1}{(E_0-H_0)'}\,V_R \Big>$            & $^3S$ & -30611.3035 \\
$S_2 $&$ \Big<V_R\,\frac{1}{(E_0-H_0)'}\,\frac{1}{r^3}\Big>$   & $^3S$ &      4.33362 \\
$S_3 $&$ \Big<V_R\,\frac{1}{(E_0-H_0)'}\,H_R \Big>$            & $^3S$ &     2399.03334 \\
$S_4 $&$ \bigg\langle\Big(\frac{\vec r_1}{r_1^3}\times\vec p_1+\frac{\vec r_2}{r_2^3}\times\vec p_2\Big)\,
\frac{1}{(E_0-H_0)'}\,
\Big(
\frac{\vec{ r}_1}{r_1^3}\times\vec{ p}_1+\frac{\vec{ r}_2}{r_2^3}\times\vec{ p}_2
\Big)\Big>  $                                    & $^3P^e$ &    -0.01580 \\
$ S_5 $&$ \bigg\langle\Big(\frac{\vec r_1}{r_1^3}\times\vec p_1+\frac{\vec r_2}{r_2^3}\times\vec p_2\Big)\,
\frac{1}{(E_0-H_0)'}\,\Big(
\frac{\vec{ r}}{r^3}\times(\vec{ p}_1-\vec{ p}_2)\Big)\Big>
 $                                               & $^3P^e$ &    -0.03892 \\
$
S_6 $&$ \Big< \Big(\frac{\delta^{ij}}{r_1^3} - \frac{3 r_1^i r_1^j}{r_1^5}
+\frac{\delta^{ij}}{r_2^3} - \frac{3 r_2^i r_2^j}{r_2^5}\Big)\,\frac{1}{(E_0-H_0)'}\,\left(
\frac{\delta^{ij}}{r^3}
-3\,\frac{r^i r^j}{r^5}\right)\Big>
 $                                               & $^3D^e$ &    -0.07872 \\
\end{tabular}
\end{ruledtabular}
\end{table*}

\subsection{Regularization of divergencies and restoration of the
$\delta^3(r_a)$ part}
\label{sec:matching}

From now on we will use atomic units and pull out the overall $\alpha^7$ prefactor.
The second-order matrix elements in Eq.~(\ref{1}) contain divergences coming from the
summation over the intermediate states. They arise when operators on the left
and on the right of the resolvent $1/(E_0-H_0)'$ are sufficiently singular, so that their first-order
matrix elements are finite but the second-order matrix elements diverge. Specifically, there
are two such ``problematic'' operators in our case, the electron-nucleus Dirac $\delta$ function and
the spin-independent part of the Breit Hamiltonian $H^{(4)}_{\rm nfs}$, given by Eq.~(6) of
Ref.~\cite{pachucki:09:hefs}. The divergences become more tractable if one moves them
to first-order matrix elements. This can be accomplished \cite{pachucki:06:hesinglet} by representing the problematic singular
operators as an anticommutator with the Schr\"odinger Hamiltonian $H_0$ plus some more regular operator.
Specifically, for the Dirac $\delta$ function, we use the following identity
\begin{align}\label{regularize}
4\pi Z\,\big[\delta^3(r_1) + \delta^3(r_2)\big] =&\ \{ H_0-E_0,Q\} + V_R\,,\\
Q = &\ 2\bigg(\frac{Z}{r_1}+\frac{Z}{r_2}\bigg)\,,\label{Q}
\end{align}
where $V_R$ is the regularized operator
defined by its action on an eigenfunction $\phi$ of Hamiltonian $H_0$ on the right as
\begin{align}\label{eq:VA}
V_R |\phi\rangle = -2\bigg(\frac{Z\vec r_1}{r_1^3}\cdot\vec\nabla_1+\frac{Z\vec r_2}{r_2^3}\cdot\vec\nabla_2\bigg)|\phi\rangle\,.
\end{align}
For the spin-independent part of the Breit Hamiltonian we use a similar identity,
\begin{align}\label{A4}
H^{(4)}_\mathrm{nfs} = &\ \{H_0-E_0,\tilde Q\} + H_R \,,\\
\tilde Q = &\ -\frac{1}{4} \bigg(\frac{Z}{r_1}+\frac{Z}{r_2}\bigg)\,,
\end{align}
where $H_R$ is defined by its action on the eigenfunction of $H_0$ on the right as
\begin{align}\label{eq:HA}
&\ H_R|\phi\rangle = \bigg[\frac14 p_1^2 p_2^2 - \frac12 (E_0-V)^2 - \frac{1}{2}\,p_1^i\bigg(\frac{\delta^{ij}}{r} + \frac{ r^i r^j}{r^3}\bigg)\,p_2^j \nonumber\\
&\ - \frac{Z}{4}\frac{\vec r_1\cdot\vec\nabla_1}{r_1^3} - \frac{Z}{4}\frac{\vec r_2\cdot\vec\nabla_2}{r_2^3}
+\frac{\vec r}{2r^3}\cdot(\vec\nabla_1-\vec\nabla_2)\bigg]|\phi\rangle\,,
\end{align}
where $V = -Z/r_1-Z/r_2+1/r$.
By applying these identities as described in Appendix~\ref{app:2ndorder}, we express the second-order
contributions in Eq.~(\ref{1}) as
\begin{align}
2\,\bigg\langle H^{(4)}_{\rm hfs}&\,\frac{1}{(E_0-H_0)'}\,H^{(5)}\bigg\rangle
 +2\,\bigg\langle H^{(5)}_{\rm hfs}\frac{1}{(E_0-H_0)'}\,H^{(4)}\bigg\rangle
 \nonumber \\ &
 = E_{\rm sec}(\mathrm{se}) + E_{\rm sec}(\mathrm{vp}) + E_{{\rm fo}, A}\,,
\end{align}
where $E_{\rm sec}(\mathrm{se})$ and $E_{\rm sec}(\mathrm{vp})$ are the finite second-order contributions 
given by Eqs.~(\ref{sofin}) and (\ref{E7})
that correspond to the self-energy and vacuum polarization, respectively, and
$E_{{\rm fo}, A}$ is an additional first-order contribution.
As previously, in our derivation we dropped terms proportional to the electron-nucleus Dirac $\delta$ function
in the first-order matrix elements, which will be restored later.

A similar regularization of the Fermi contact interaction was carried out in the calculation of the
low-energy part $E_L$, for the integrand $P_L(k)$ in Eq.~(\ref{eq:PL}). The integrand $P_L(k)$ does not
contain any divergences, so the regularization is not obligatory but it greatly improves the
convergence of numerical basis-set calculations. For the first term in the right-hand side of Eq.~(\ref{eq:PL}) we
used the representation (\ref{regularize}), whereas for the second term we employed a more general identity
\begin{align}\label{regularize2}
4\pi Z\,\big[\delta^3(r_1) + \delta^3(r_2)\big] =
\big\{H_0-E_0,Q\big\} + \widetilde{V}_R
\,,
\end{align}
where
\begin{align}\label{regularize2b}
\widetilde{V}_R
 &\, = 4\,(E_0-V)\bigg(\frac{Z}{r_1}+\frac{Z}{r_2}\bigg)
  \nonumber \\ &\
- 2\,\vec p_1\bigg(\frac{Z}{r_1}+\frac{Z}{r_2}\bigg)\,\vec p_1
- 2\,\vec p_2\bigg(\frac{Z}{r_1}+\frac{Z}{r_2}\bigg)\,\vec p_2
\,,
\end{align}
and $Q$ is defined in Eq.~(\ref{Q}). It might be noted that
the dependence on $E_0$ in the above equations cancels out, so that they represent
a general operator identity.

Now we turn to restoring the contribution proportional to the electron-nucleus
Dirac $\delta$ function. This is accomplished by evaluating the large-$Z$ limit of
the derived $m\alpha^7$ contributions. In the $Z$$\to$$\infty$ limit, all effects
of the electron-electron interaction vanish (since they are suppressed by a factor of $1/Z$
as compared to the electron-nucleus interaction) and the result should agree
with the $m\alpha^7$ correction derived for the hydrogenlike ions. This
matching gives us the coefficient at the electron-nucleus
Dirac $\delta$ function. The evaluation of the large-$Z$ limit of our formulas and the
matching with the hydrogenic results is described
in Appendix \ref{app:hydrSE}. As a result, we obtain an additional first-order
contribution proportional to the electron-nucleus
Dirac $\delta$ function,
\begin{align}\label{eq:eta}
E_{{\rm fo}, B} = \eta\,\lbr \vec{I}\cdot\vec{S}\rbr\, Z^3\pi
 \lbr \big[ \delta^3(r_1)+\delta^3(r_2)\big] \rbr\,,
\end{align}
with the coefficient $\eta$ given by
\begin{align}\label{eq:eta2}
\eta = &\ \frac{g}{4\,\pi\,M}
\bigg[-\frac{5351}{1350} - \frac{44\pi^2}{27} - \frac{10}{3}\zeta(3) + \frac{896}{27}\ln 2\nonumber\\
&\ + \frac{16}{9}\ln^2 2  - \frac{4882}{135}\ln\alpha - \frac{64}{9}\ln^2\alpha + \frac{256}{9}\ln 2\ln\alpha \bigg]\,,
\end{align}
where we dropped the $\lambda$-dependent terms.

Finally we obtain the total first-order contribution as
\begin{align}
\lbr H^{(7)}_{ {\rm hfs},A}\rbr + \lbr H^{(7)}_{ {\rm hfs},B}\rbr +  E_{{\rm fo}, A} + E_{{\rm fo}, B}
= E_{{\rm fo}}(\mathrm{se}) + E_{{\rm fo}}(\mathrm{vp}) \,,
\end{align}
where $E_{{\rm fo}}(\mathrm{se})$ and $E_{{\rm fo}}(\mathrm{vp})$ are given by Eqs.~(\ref{E3}) and (\ref{E8}), respectively.

So far the individual $m\alpha^7$ contributions depend on the
logarithm of the low-energy cutoff of photon momenta, $\ln\lambda$. Naturally, the
complete  $m\alpha^7$ correction should not depend on $\lambda$. The cancellation
of the $\lambda$-dependent terms is demonstrated in Appendix~\ref{app:lambda}; this
constituted an important cross-check of the derivation. After the cancellation is
proven, we set $\lambda \to 1$ in all formulas.

\subsection{Final formulas}

We now collect all the contributions and obtain the complete result for
the $m\alpha^7$ hfs correction for the ${}^3 S$ states of heliumlike ions.
It is convenient to separate out the dependence on the total angular momentum and
write the final result as
\begin{align}
E^{(7)} 
= &\ \big<\vec I\cdot\vec S\big> \, A^{(7)}\,,
\end{align}
where
\begin{equation}
A^{(7)} = A_L + A_{\rm fo}(\mathrm{se}) + A_{\rm sec}(\mathrm{se}) + A_{\rm fo}(\mathrm{vp}) 
 + A_{\rm sec}(\mathrm{vp})  + A_{\rm 2loop}. 
\end{equation}
Here,
the first three terms come from the one-loop
self-energy, the fourth and fifth terms are the one-loop
vacuum-polarization contribution, and the last term is the two-loop correction.
The low-energy self-energy contribution $E'_L = \big<\vec I\cdot\vec S\big>\,A_L $ is given by Eq.~(\ref{EL}).
The first-order self-energy contribution $A_{\rm fo}$ is conveniently expressed in terms of
$Q_i$ operators which were encountered in our previous investigation
of the $m\alpha^7$ effects to the Lamb shift \cite{patkos:21:helamb} and are defined in Table~\ref{tab:oprsQ1}.
The result is
\begin{widetext}
\begin{align}\label{fofinQ}
A_{\rm fo}(\mathrm{se})
  = &\ \frac{g}{2\,\pi\,M}\bigg\{
  \frac{1}{9}\, \bigg(\frac{71}{3}+32\ln\frac{\alpha^{-2}}{2}\bigg)\, Z^2\, Q_1\,Q_{53}
 +\bigg(\frac{143}{108} + \frac89\ln\frac{\alpha^{-2}}{2}\bigg)\,Z^2\, Q_{57}
 \nonumber\\
&\ -\frac{1}{3}\, \bigg(\frac{85}{6}+16\ln\frac{\alpha^{-2}}{2} \bigg) \,\frac{Z^2}{2}\, Q_3
 - \frac{56}{9}\,Z\,Q_9\,Q_{53}
 +\frac{56}{9}\,Z\, Q_{59}
-\frac{13}{12}\,Z\,Q_{18}
 + \frac{4Z}{3}\, E^{(4)}\,Q_{53}
\nonumber\\&\
 +\frac{2Z}{3} \Big(
-2 E_0 Q_{13} + Q_{17} + E_0^2 Q_{53} + 2 Z E_0 Q_{11} + 2 Z E_0 Q_{12} - 2 Z Q_{14} - 2 Z Q_{16} + 3 Z^2 Q_{15} +  Z^2 Q_{56}
 \Big)
\nonumber\\&\
 -\frac{Z}{3}\,Q_{28}
 + \frac{2Z}{3}\,Q_{24}
+\frac{Z}{36}\bigg(\frac{77}{6} + 16 \ln\frac{\alpha^{-2}}{2} \bigg)\,Q_{51}
-\frac{Z}{36}\bigg(\frac{95}{3} + 32 \ln\frac{\alpha^{-2}}{2} \bigg)
\Big( E_0\,Q_1 - Q_3 -\frac12 \,Q_4 \Big)
\nonumber\\
&\ + \bigg[-\frac76 - \frac{44\pi^2}{27} - \frac{10}{3}\zeta(3) + \frac{896}{27}\ln 2+ \frac{16}{9}\ln^2 2
- \frac{938}{27}\ln\alpha - \frac{64}{9}\ln^2\alpha + \frac{256}{9}\ln 2\ln\alpha \bigg]
\,\frac{Z^3}{4}\,Q_1\bigg\}
\,.
\end{align}
The second-order self-energy contribution is
\begin{align}\label{sofin2}
A_{\rm sec}(\mathrm{se}) = &\ \frac{g}{2\,\pi\,M}\bigg\{
\frac{2}{9}\,
\bigg[\bigg(\frac{5}{6}+\ln\frac{\alpha^{-2}}{2}\bigg)\,S_1 - 7\,S_2 + \frac32\, S_3\bigg]
 + \frac{Z}{3}\,
\bigg(\frac{Z}{2}\,S_4 - S_5\bigg)
 -\frac{Z}{8}\,S_6\bigg\}
\,,
\end{align}
\end{widetext}
where the second-order matrix elements $S_i$ are defined in Table~\ref{table:sec}.
The vacuum-polarization contribution is given by
\begin{align}\label{finvp2}
 A_{\rm fo}(\mathrm{vp}) = &
-\frac{g}{45\,\pi\,M}
\bigg[ 16 Z^2 \, Q_1\,Q_{53}
+ 2 Z\, Q_{51}
+4Z^2 Q_{57}
\nonumber\\ &
+4Z(1-3Z)\, Q_3
- 4 Z\,\Big( E_0\,Q_1 -\frac12 Q_4 \Big)
\nonumber
\\ &
+ Z^3\,\bigg(\frac{236}{15} + 8\ln\alpha\bigg)\, Q_1
\bigg]\,,
\end{align}
and
\begin{align}
 A_{\rm sec}(\mathrm{vp}) = &\
-\frac{g}{45\,\pi\,M}
S_1\,.
\end{align}
The two-loop QED contribution $E_{\rm 2loop} = \big<\vec I\cdot\vec S\big>\,A_{\rm 2loop} $ is given by Eq.~(\ref{2loop}).

%
\section{Results}

\begin{table}
\caption{Individual $m\alpha^7$ corrections to the magnetic dipole
hfs of the $2^3S$ state in Li$^+$. Units are $\alpha^3\,A_F$.
\label{table:E7}
}
\begin{ruledtabular}
\begin{tabular}{lw{5.8}}
\multicolumn{1}{l}{Term} &
        \multicolumn{1}{c}{Value}
\\
\hline\\[-5pt]
 ${ A}_L$               &       45.0968\,(22)  \\
 ${ A}_\mathrm{fo}(\mathrm{se})$       &       50.8070   \\
 ${ A}_\mathrm{sec}(\mathrm{se})$      &     -186.2134    \\
 ${ A}_\mathrm{fo}(\mathrm{vp})$       &        1.3752 \\
  ${ A}_\mathrm{sec}(\mathrm{vp})$       &      3.7756   \\
  ${ A}_\mathrm{2loop}$   &         0.7362 \\[1ex]
 ${ A}^{(7)}$             &      -84.4226\,(22)  \\
\end{tabular}
\end{ruledtabular}
\end{table}

\subsection{Li$^+$ hfs}

Our numerical calculations of the $m\alpha^7$ corrections
were carried out with the basis set of exponential functions
$e^{-\alpha_i\,r_1-\beta_i\,r_2-\gamma_i\,r}$ introduced by Korobov \cite{korobov:00}.
The method of calculations follows the one developed in our previous investigations
and reviewed in Ref.~\cite{yerokhin:21:hereview}. The most difficult numerical part
is the computation of the Bethe-logarithm contribution $E_L$. This contribution
is very similar to the low-energy $m\alpha^7$ contributions for the Lamb shift;
so we refer the reader to our previous work~\cite{yerokhin:18:betherel}
for details of the numerical approach.
Expressed in terms of $\beta_{\delta}$, our numerical results for the low-energy
$m\alpha^7$ contribution are
\begin{align}
\beta_{\delta}(2^3S, Z = 2) = 70.5314\,(28)\,, \\
\beta_{\delta}(2^3S, Z = 3) = 70.0036\,(34)\,,
\end{align}
which can be compared with the hydrogenic limit \cite{jentschura:03:jpa}
\begin{align}
\beta_{\delta}(1s2s, Z=\infty) = 68.834\,482\,.
\end{align}
Numerical results for the individual $m\alpha^7$ corrections to the hfs of the $2^3S$ state
of Li$^+$ are presented in Table~\ref{table:E7}.

We now collect all available theoretical contributions to the magnetic dipole hfs in $^6$Li$^+$
and $^7$Li$^+$. Accurate values of the nuclear magnetic moments
were obtained in Ref.~\cite{pachucki:unpunlished},
\begin{align}\label{eq:mu}
\frac{\mu}{\mu_N} =
   \left\{
   \begin{array}{ll}
   0.822\,044\,63\,(37)  & \mbox{\rm for}\ \  ^6\mathrm{Li}\,,\\
   3.256\,416\,19\,(55)  & \mbox{\rm for}\ \  ^7\mathrm{Li}\,,
     \end{array}\right.
\end{align}
and, therefore,
\begin{align}
g(^6\mathrm{Li}) = &\ 1.635\,878\,84(74)\,, \nonumber \\
g(^7\mathrm{Li}) =&\ 5.039\,258\,37(85)\,.
\end{align}
The nuclear masses
\begin{align}
M({}^6\mathrm{Li}) =&\  6.013\,477\,3618(15)\,\mathrm{u},\nonumber \\
M({}^7\mathrm{Li}) =&\  7.014\,357\,9087(45)\, \mathrm{u},
\end{align}
were obtained from the atomic masses from Ref.~\cite{wang:21} by subtracting
the electron rest masses and the binding energies.
Values of other physical constants were taken from Ref.~\cite{tiesinga:21:codata18}.

Table~\ref{table:liplus} presents results for individual theoretical contributions
to the magnetic dipole hfs of the $2^3S_1$ state in
${}^6\mathrm{Li^+}$ and ${}^7\mathrm{Li}^+$.
The numerical results are
expressed in terms of $\delta$ defined by Eq.~(\ref{eq:delta}).
The theoretical uncertainty is defined by the QED contribution of order $m\alpha^8$,
for which no direct calculations exist so far; it was estimated
by using the corresponding hydrogenic result listed in Sec.~\ref{sec:hydr}.
The entry $\delta_{\rm theo}$ is the total theoretical prediction without
the nuclear-structure contribution. The difference $\delta_{\rm exp}-\delta_{\rm theo}$ 
then determines the nuclear-structure contribution $\delta_{\rm struct}$.

\begin{table}[t]
\caption{Contributions to the magnetic dipole hfs of the $2^3S_1$ state in
${}^6\mathrm{Li^+}$ and ${}^7\mathrm{Li}^+$ and the determination of
the nuclear structure contribution $\delta_\mathrm{struct}$.
\label{table:liplus}
}
\begin{ruledtabular}
\begin{tabular}{lw{1.10}w{1.10}}
\multicolumn{1}{l}{Term} &
        \multicolumn{1}{c}{${}^6\mathrm{Li}^+$} &
                    \multicolumn{1}{c}{${}^7\mathrm{Li}^+$}
\\
\hline\\[-5pt]
 $\kappa$  & 0.001\,159\,7  & 0.001\,159\,7 \\
 $\delta^{(2)}$ & 0.000\,443\,5 & 0.000\,443\,5    \\
 $\delta^{(3)}$ &-0.000\,032\,8 &  -0.000\,032\,8    \\
 $\delta^{(4)}$ &-0.000\,002\,1(5)   &  -0.000\,002\,1(5)     \\
 $\delta_\mathrm{rec.mix}^{(2)}$ & 0.000\,002\,4   & 0.000\,006\,2    \\
 $\delta_\mathrm{rec.rel}^{(2)}$  &  0.000\,000\,3  &   0.000\,000\,4  \\
 $\delta_\mathrm{theo}$               & 0.001\, 570\, 9(5) & 0.001\, 574\, 9(5) \\ 
 $\delta_\mathrm{theo}$ \cite{qi:20}  & 0.001\, 576(2)     & 0.001\, 580(2) \\[1ex]
 $A_F$[GHz] &2.997\,908\,1(14) & 7.917\,508\,1(13) \\
 $A_\mathrm{exp}$[GHz] \cite{sun:23, guan:20} & 3.001\,805\,1(7) & 7.926\,990\,1(23) \\
 $\delta_\mathrm{exp}=A_\mathrm{exp}/A_F-1$      & 0.001\,299\,9(24) & 0.001\,197\,6(29) \\ [1ex]
 $\delta_\mathrm{struct} =\delta_\mathrm{exp} - \delta_\mathrm{theo}$ & -0.000\,271\,0(24) & -0.000\, 377\, 3(30)
\end{tabular}
\end{ruledtabular}
\end{table}

\subsection{Effective Zemach radius}
The nuclear structure contribution $\delta_{\rm struct}$ is parameterized in terms of the effective Zemach
radius $\tilde{r}_Z$ according to Eq.~(\ref{eq:rZtilde}). Numerical results for $\tilde{r}_Z$ of
$^{6,7}$Li are listed in Table~\ref{table:Zemach}.
This table also compares the present values of $\tilde{r}_Z$ with previous
determinations. The result from Puchalski {\em et al.} \cite{puchalski:13} 
was recalculated by including $\delta_\mathrm{rec.rel}$
and by using the updated magnetic moments of $^{6,7}$Li, given by Eq.~(\ref{eq:mu}).
We confirm the surprising result,
pointed out in Ref. \cite{puchalski:13}, that the effective Zemach radius of $^6$Li
is smaller than for $^7$Li, in spite of the fact that the nuclear charge radius of $^6$Li
is larger than for $^7$Li. The probable explanation is a large contribution of
inelastic effects. Previously, significant inelastic contributions were found in
hfs of D \cite{friar:05} and $\mu$D \cite{kalinowski:18}.

\begin{table}[t]
\caption{Results for the effective Zemach radius $\tilde{r}_Z$ of $^6$Li and $^7$Li, in fermi.
}
\label{table:Zemach}
\begin{ruledtabular}
\begin{tabular}{llw{2.6}w{2.6}}
\multicolumn{1}{l}{System} &
    \multicolumn{1}{l}{Reference} &
        \multicolumn{1}{c}{${}^6\mathrm{Li}$} &
                    \multicolumn{1}{c}{${}^7\mathrm{Li}$}
\\
\hline\\[-5pt]
      Li$^+$  & This work                                   &  2.39\,(2)  &  3.33\,(3)\\
      Li$^+$  & Sun {\em et al.} \cite{sun:23}              &  2.44\,(2)              \\
      Li$^+$  & Qi {\em et al.} \cite{qi:20}                &  2.40\,(16) &3.33\,(7) \\
      Li$^+$  & Qi {\em et al.} \cite{qi:20}                &  2.47\,(8)  & 3.38\,(3) \\
      Li         & Puchalski {\em et al.} \cite{puchalski:13}$^{\dag}$  &  2.29\,(4)  & 3.23\,(4)
\end{tabular}
$^{\dag}$ recalculated for the nuclear momenta given by Eq.~(\ref{eq:mu}).
\end{ruledtabular}
\end{table}

\subsection{Li$^{2+}$ hfs}

Theory of the magnetic dipole hfs of hydrogenlike atoms is summarized by Eqs.~(\ref{eq:13})-(\ref{10}).
This theory alone is not capable of predicting the hfs energy splittings since
the nuclear structure contribution
$\delta_{\rm struct}$ cannot be accurately calculated from the first principles
at present. We can circumvent this problem by using the nuclear structure contribution
extracted from the Li$^+$ hfs measurements in order to predict the
Li$^{2+}$ hfs. An equivalent way is to calculate the difference of the normalized hfs values in Li$^{2+}$ and
Li$^+$ and use the experimental result for the Li$^+$ hfs to predict the hfs in Li$^{2+}$.
Such a determination is presented in Table~\ref{table:hydrlike}. The table lists theoretical
values for $\delta^{(2)}$, $\delta^{(2)}_\mathrm{rec.mix}$, and $\delta^{(3)}$
for the $\mbox{\rm Li}^{2+}$-$\mbox{\rm Li}^+$ difference.
The sum of the theoretical contributions and the experimental value $\delta_\mathrm{exp}(\mbox{\rm Li}^{+})$
gives the prediction for $\delta(\mbox{\rm Li}^{2+})$.
It is remarkable that
the uncertainty of our prediction for the Li$^{2+}$ hfs comes exclusively from the uncertainty
of the experimental Li$^+$ hfs value.

\begin{table}
\caption{Hfs splitting in Li$^{2+}$.
}
\label{table:hydrlike}
\renewcommand{\arraystretch}{1.2}
\begin{ruledtabular}
\begin{tabular}{lw{2.12}w{3.12}}
\multicolumn{1}{l}{Term} &
        \multicolumn{1}{c}{${}^6\mathrm{Li}$} &
        \multicolumn{1}{c}{${}^7\mathrm{Li}$}
\\
\hline\\[-5pt]
  $\delta^{(2)}(\mbox{\rm Li$^{2+}$-Li$^+$})$ & -0.000\,013\,3  &  -0.000\,013\,3     \\
  $\delta^{(2)}_\mathrm{rec.mix}(\mbox{\rm Li$^{2+}$-Li$^+$})$ & -0.000\,002\,4  & -0.000\,006\,2 \\
  $\delta^{(3)}(\mbox{\rm Li$^{2+}$-Li$^+$})$ & -0.000\,000\,5  &  -0.000\,000\,5 \\
  $\delta_\mathrm{exp}(\mbox{\rm Li}^+)$ \cite{sun:23, guan:20} & 0.001\,299\,9\,(24)  &   0.001\,197\,6\,(29)  \\
  $\delta(\mbox{\rm Li}^{2+})$   & 0.001\,283\,8\,(24) &   0.001\,177\,6\,(29)  \\ [1ex]
  $E_F(\mbox{\rm Li}^{2+})\,$[GHz] & 8.468\,319(4)    & 29.819\,898(5)  \\
  $E_{\rm hfs}(\mbox{\rm Li}^{2+})\,$[GHz]    & 8.479\,190\,(21) & 29.855\,013\,(86) \\
\end{tabular}
\end{ruledtabular}
\end{table}

\subsection{Li$-$Li$^{+}$ hfs difference}

Our present calculation of hfs in Li$^+$ allows us
to check the consistency between the theoretical and experimental results
for hfs in Li and Li$^+$.
Only a few theoretical contributions to $\delta(\mbox{\rm Li-Li$^+$})$ are nonvanishing,
namely, the relativistic and QED terms $\delta^{(2)}$ and $\delta^{(3)}$,
and the hyperfine mixing contribution $\delta^{(2)}_\mathrm{rec.mix}$ (absent in the case of the Li atom).
The results are presented in Table \ref{table:diff}, where we used the Li result for $\delta^{(2)}$
from Ref.~\cite{puchalski:13}. The dominant theoretical uncertainty comes from the
estimation of the $\delta^{(3)}$ correction for Li, which we assumed to be the same as in Li$^{2+}$.
We observe a
$2\,\sigma$ tension between the theoretical and experimental hfs results, which might
result from a larger than expected $\delta^{(3)}$ correction in atomic Li.
This supposition can be verified by a direct calculation of this QED correction  in atomic Li.

\begin{table}
\caption{Li$-$Li$^+$ hfs difference.}
\label{table:diff}
\renewcommand{\arraystretch}{1.2}
\begin{ruledtabular}
\begin{tabular}{lw{2.12}w{2.12}}
\multicolumn{1}{l}{Term} &
        \multicolumn{1}{c}{${}^6\mathrm{Li}$} &
        \multicolumn{1}{c}{${}^7\mathrm{Li}$}
\\
\hline\\[-5pt]
  $\delta^{(2)}(\mbox{\rm Li-Li}^+)$ & 0.000\,204\,8  & 0.000\,204\,8      \\
  $\delta^{(2)}_\mathrm{rec.mix}(\mbox{\rm Li-Li}^+)$ & -0.000\,002\,4  & -0.000\,006\,2 \\
  $\delta^{(3)}(\mbox{\rm Li-Li}^+)$ & -0.000\,000\,5\,(47)  & -0.000\,000\,5\,(47)  \\[1ex]
  $\delta(\mbox{\rm Li-Li}^+)_\mathrm{theo}$   & 0.000\,201\,9\,(47) &   0.000\,198\,0\,(47)  \\
  $\delta(\mbox{\rm Li-Li}^+)_\mathrm{exp}$ \cite{sun:23,guan:20,beckmann:74}  & 0.000\,212\,9\,(24)  &  0.000\,209\,5\,(29) \\
\end{tabular}
\end{ruledtabular}
\end{table}

\subsection{$^6$Li-$^7$Li isotopic hfs difference}
A further test of consistency of the measured hfs values can be obtained
by examining the isotope shift of the normalized hfs values in Li and Li$^+$. On the
theoretical side, all QED contributions vanish in the isotope-shift difference. The
only noticeable correction is the nuclear recoil contribution, which is nevertheless
tiny and amounts to $\delta^{(2)}_{\rm rec}(\mbox{$^6$Li-$^7$Li}) =
-1\times 10^{-7}$. Therefore,
the $^6$Li-$^7$Li difference of the normalized experimental hfs values
can be almost solely attributed to the nuclear structure effect.
This means that the isotope-shift difference of the nuclear structure contributions can be
extracted from the experimental hfs values of atomic Li
almost without any theoretical input, see Table~\ref{table:isoshift}.
The table presents the $^6$Li-$^7$Li isotope shift  of the nuclear-structure contribution
$\delta_\mathrm{struc}$
obtained from the experimental hfs values of atomic Li \cite{beckmann:74}.
The result is compared with the
corresponding value extracted from Li$^+$ (see Table~\ref{table:liplus}).
We observe very good agreement of the isotope-shift differences of
$\delta_\mathrm{struc}$ obtained from atomic Li and Li$^+$,
which indicates the consistency of the experimental results.

\begin{table}
\caption{$^6$Li-$^7$Li isotope shift of atomic ground-state hfs.
}
\label{table:isoshift}
\begin{ruledtabular}
\begin{tabular}{lw{0.12}}
\multicolumn{1}{l}{Term} &
        \multicolumn{1}{c}{Value}
\\
\hline\\[-5pt]
  $\delta(\mbox{${}^6$\rm{Li}-${}^7$\rm{Li}})_\mathrm{theo}$ & -0.000\,000\,1   \\
  $\delta(\mbox{${}^6$\rm{Li}-${}^7$\rm{Li}})_\mathrm{exp}$  \cite{beckmann:74}   & 0.000\,105\,7(5)  \\[1ex]
  $\delta_\mathrm{struc}(\mbox{${}^6$\rm{Li}-${}^7$\rm{Li}})$ & 0.000\,105\,8(5) \\
  $\delta_\mathrm{struc}(\mbox{${}^6$\rm{Li}${}^+$-${}^7$\rm{Li}${}^+$})$ & 0.000\,106\,3(38) \\
\end{tabular}
\end{ruledtabular}
\end{table}

This consistency can be studied further by constructing the difference $\Delta$ from the normalized hfs isotope
shifts as follows,
\begin{align}
\Delta =&\ [\delta(^6\mathrm{Li}^+) - \delta(^7\mathrm{Li}^+)] -  [\delta(^6\mathrm{Li}) - \delta(^7\mathrm{Li})] \,.
\end{align}
From the theoretical point of view, this difference
comes mostly from $\delta^{(2)}_\mathrm{rec.mix}$,
is very small numerically and can be calculated very accurately. We obtain
\begin{align}
\Delta_{\rm theo} = -3.9 \times  10^{-6}\,,
\end{align}
which can be
compared to the experimental value
\begin{align}
 \Delta_{\rm exp} = -3.4\,(3.8)\times  10^{-6}\,,
 \end{align}
obtained from the experimental results \cite{sun:23,guan:20,beckmann:74}.
The conservative estimate of uncertainty for $\Delta_{\rm exp}$ is obtained by adding quadratically
the uncertainties of individual independent
measurements, with the assumption that they are not correlated.

The above comparison constitutes a strict test of consistency of
the four different measurements of hfs in $^{6,7}$Li$^{+}$ and $^{6,7}$Li \cite{sun:23,guan:20,beckmann:74}.
By contrast,
if we use the recent value for the $^6$Li hfs: $228.201\,5(14)$ MHz from Ref.~\cite{li:20}, then
the experimental difference moves away from the theoretical prediction
and becomes
$\Delta_\mathrm{exp} = 13.1\,(7.2)\times 10^{-6}$, which casts some doubts about the correctness
of the uncertainty estimation in Ref.~\cite{li:20}.

\section{Conclusion}

We have performed calculations of the QED effects of order $m\alpha^7 ( = \alpha^3E_F)$ to
the magnetic dipole hyperfine structure in Li$^+$. This calculation greatly improves the
theoretical value of hfs in Li$^+$ in the point-nucleus limit. By comparing the theoretical
point-nucleus result with the experimental Li$^+$ hfs value, we determine the nuclear-structure
contribution and parametrize it in terms of the effective Zemach radius.
We confirm the surprising result,
pointed out in Ref. \cite{puchalski:13}, that the effective Zemach radius of $^6$Li
is smaller than that of $^7$Li, which is in contrast
with the charge radius of $^6$Li being
larger than that of $^7$Li. The probable explanation of this fact is large contributions of
inelastic effects, for which no calculations exist up to now.

It is demonstrated that the nuclear-structure contribution, when normalized by the
Fermi energy $E_F$, is nearly the same numerically in Li$^{2+}$, Li$^{+}$, and atomic Li.
The charge-state dependent contributions to $\delta_{\rm struct}$ are of order 
$O(Z\,\alpha)^2$ and very small numerically.
Using this statement, we obtain accurate predictions for the hfs in $^6$Li$^{2+}$
and $^7$Li$^{2+}$, for which no experimental data is available so far.
Examination of the normalized differences of the hfs values of Li$^{+}$ and Li and
of the corresponding isotope-shift differences allowed us to demonstrate the consistency
of four different measurements of hfs in $^{6,7}$Li$^{+}$ and $^{6,7}$Li
\cite{sun:23,guan:20,beckmann:74}.
By contrast, the recent measurement of  $^6$Li hfs \cite{li:20} leads to a 2$\sigma$
tension in the consistency test.

\begin{acknowledgments}
K.P. and V.P. acknowledge support from the National Science Center (Poland) Grant No. 2017/27/B/ST2/02459.
\end{acknowledgments}


\appendix

\section{The asymptotic expansion coefficients of the Bethe-logarithm contribution}\label{app:asympt}

Here we present a summary of formulas derived for the coefficients in the asymptotic expansion of
the integrand $P_L(k)$ of the low-energy contribution $E_L$ given by Eq.~(\ref{eq:as}).
As in Sec.~\ref{lowen}, we work in atomic units and pull out the overall $\alpha^7$ prefactor.
The asymptotic coefficient $A$ is
\begin{equation}\label{coef:A}
A = -2\, \bigg\langle \vec P^2\,\frac{1}{(E_0-H_0)'}\,{\cal V}_F\bigg\rangle
\end{equation}
where $\vec P = \vec p_1+\vec p_2$ and ${\cal V}_F$ is given by Eq.~(\ref{eq:VF}).
For the numerical evaluation, we transform the second-order
matrix element to a more regular form by using the identity (\ref{regularize}).
After simple calculation, we obtain
\begin{align}
A = &\ \frac{g}{3\,M}\,\langle\vec I\cdot\vec S\rangle
\bigg[-\bigg\langle \vec P^2\,\frac{1}{(E_0-H_0)'}\,V_R\bigg\rangle \nonumber\\
&\ + 4\,\pi\,Z \,\langle\delta^3(r_1)+\delta^3(r_2)\rangle
+2\,\bigg\langle \vec P\,\bigg(\frac{Z}{r_1}+\frac{Z}{r_2}\bigg)\,\vec P\bigg\rangle\nonumber\\
&\ + \langle\vec P^2\rangle\bigg(4 E_0 - \bigg\langle \frac{2}{r}\bigg\rangle\bigg)\bigg]\,.
\end{align}

The coefficients $B$ and $C$ originate from the exchange of high-momenta photons.
The corresponding formulas are derived by considering
the forward scattering amplitude with two and three photon exchanges, correspondingly,
perturbed by the Fermi contact interaction.
The results are proportional to the expectation values of
the local contact interaction and are given by
\begin{align}
B =&\  -\frac{8\sqrt{2}\,g\,Z^2}{3\,M}\,\langle\vec I\cdot\vec S\rangle\,\langle \pi\big(\delta^3(r_1)+\delta^3(r_2)\big)\rangle\,,
 \\
C =&\  -\frac{4\,g\,Z^3}{3\,M}\,\langle\vec I\cdot\vec S\rangle\,\langle \pi\big(\delta^3(r_1)+\delta^3(r_2)\big)\rangle\,.
\end{align}

The calculation of the coefficient $D$ is more complicated. It consists of the low- and high-energy parts
which are calculated separately using the dimensional regularization, similar to that for the Lamb shift \cite{yerokhin:18:betherel}.
The result is
\begin{align}
D = &\ \frac{g}{6\,M}\,\langle\vec I\cdot\vec S\rangle\bigg[
\langle 16\pi\,Z^2\big(\delta^3(r_1)+\delta^3(r_2)\big)\rangle\bigg\langle\frac{1}{r_1}+\frac{1}{r_2}\bigg\rangle \nonumber\\
&\ +2\,\bigg\langle \frac{Z^2}{r_1^4}+\frac{Z^2}{r_2^4}\bigg\rangle
+\bigg\langle V_R\,\frac{1}{(E_0-H_0)'}\,V_R\bigg\rangle\nonumber\\
&\ + \bigg<
 \vec p_1\,4\pi\,Z\,\delta^3(r_1)\,\vec p_1  - \bigg(E_0+\frac{(3Z-1)}{r_2} - \frac{p_2^2}{2} \nonumber \\
&\  - 6 Z^2 + 5 Z^2\ln2\bigg)
\,8\pi\,Z\,\delta^3(r_1) + (1\leftrightarrow2)
 \bigg>\bigg].
\end{align}

\section{Derivation of $H^{(7)}_{\rm hfs}$}\label{app:H7}

In this section we derive $H^{(7)}_{ {\rm hfs},A}$ and $H^{(7)}_{ {\rm hfs},B}$ in Eq.~(\ref{H7se}).
The effective operator $H^{(7)}_{ {\rm hfs},A}$ is given by Eq.~(\ref{HBP}) which can be rewritten as
\begin{align}\label{H7a1}
&\ H^{(7)}_{ {\rm hfs},A}
=
\sum_a \kappa \bigg[\frac{Z\alpha}{2}\, \vec\sigma_a\cdot \frac{\vec r_a}{r_a^3}\times \big(-e\vec A_a\big)\nonumber\\
&\ - \frac{e}{16} \, \vec\sigma_a\cdot\Delta\vec B_a
+\frac{e}{4}\,(\vec p_a\cdot \vec\sigma_a) (\vec B_a \cdot\vec p_a)\bigg]
\nonumber\\&\
+\sum_{a,b;a\neq b}\kappa\,\frac{\alpha}{2 r_{ab}^3}\, \vec\sigma_a\cdot\vec r_{ab}\times\big(e\vec A_a- e\vec A_b \big)\,.
\end{align}
The individual parts of this expression
are calculated as follows,
\begin{widetext}
\begin{align}
\sum_a \frac{Z\alpha}{2 }\, \vec\sigma_a\cdot \frac{\vec r_a}{r_a^3}\times \big[-e\vec A(\vec r_a)\big]
= &\ \frac{g\,(Z\alpha)^2}{3\,M}\sum_a\frac{\vec s_a\cdot\vec I}{r_a^4} \,,\\
\sum_a \frac{e}{4}\,(\vec p_a\cdot \vec\sigma_a) (\vec B(\vec r_a) \cdot\vec p_a) =&\
\frac{g\,Z\alpha}{12\,M}\sum_a \vec s_a \cdot \vec I\,\bigg[-\frac{8\pi}{3}\,p_a^i\,\delta^3(r_a)\,p_a^i
+p_a^i\,\frac{1}{r_a^3}\bigg(\delta^{ij}-3\frac{r_a^i r_a^j}{r_a^2}\bigg)\,p_a^j\bigg]\,,\\
\sum_a - \frac{e}{16}\,\vec\sigma_a\cdot\,\Delta\vec B(\vec r_a) =&\
\frac{g\,Z\alpha}{6\,M}\sum_a  \vec s_a\cdot\vec I\,\Delta\,\pi\,\delta^3(r_a)\,,\\
\sum_{a,b;a\neq b}\frac{\alpha}{2 r_{ab}^3}\, \vec\sigma_a\cdot\vec r_{ab}\times\big(e\vec A(\vec r_a)- e\vec A(\vec r_b) \big)
= &\ -\frac{g\,Z\alpha^2}{3\,M}\sum_{a,b;a\neq b} \vec s_a\cdot \vec I\,\frac{\vec r_{ab}}{r_{ab}^3}
\,\cdot\bigg(\frac{\vec r_a}{r_a^3}-\frac{\vec r_b}{r_b^3}\bigg)\,.
\end{align}
\end{widetext}
Deriving the above formulas we used $\langle S^i I^j\rangle = \langle \vec I\cdot\vec S\rangle\,\delta^{ij}/3$, 
which is valid for the triplet $S$ states. After simple calculations
we obtain the result for $H^{(7)}_{ {\rm hfs},A}$ as given by Eq.~(\ref{H7a}).

Turning to $H^{(7)}_{ {\rm hfs},B}$, we evaluate it as
\begin{align}
H^{(7)}_{ {\rm hfs},B} = &\
 \sum_a -\frac{e}{2} \, \bigg[F'_1(0)+F'_2(0) - \frac{\alpha}{15\,\pi}\bigg]\,\vec\sigma_a \cdot\Delta \, \vec B(\vec r_a) \nonumber\\
= &\ \sum_a \frac{g}{2\,M}\, Z\alpha\,\bigg[F'_1(0)+F'_2(0) - \frac{\alpha}{15\,\pi}\bigg]\frac{8\pi}{3}\,\vec s_a\cdot\vec I\nonumber\\
&\ \times\Delta\,\delta^3(r_a)\,,
\end{align}
where the index $a = 1,2$ runs over the two electrons. Deriving the above formula, we
again omitted terms vanishing for the triplet $S$ states.
We simplify the result 
by rewriting all terms as a sum of symmetric and antisymmetric in spin parts. For example,
\begin{align}\label{D9}
&\ \sum_a \vec s_a\cdot\vec I\,\delta^3(r_a)
 =  \frac12\big(\vec s_1 + \vec s_2\big)\cdot\vec I\,\big[\delta^3(r_1)+\delta^3(r_2)\big]\nonumber\\
&\ +\frac12\big(\vec s_1 - \vec s_2\big)\cdot\vec I\,\big[\delta^3(r_1)-\delta^3(r_2)\big]\,.
\end{align}
The expectation value of the antisymmetric part vanishes for the triplet $S$ states
and only the symmetric part contributes.
Taking into account that the sum of the spins of both electrons is equal to the 
total spin $\vec S$, we can make the replacement
$\vec s_a\rightarrow\vec S/2$ in all expressions.

\section{Regularized form of the second-order contribution}
\label{app:2ndorder}

Here we consider the singular part of the second-order contributions in Eq.~(\ref{1}),
namely, terms containing the Dirac $\delta$ functions.
With the help of Eqs.~(\ref{regularize}) and (\ref{A4})
the singular second-order contributions are transformed
into a form suitable for numerical calculation, whereas all divergences are
transferred to the first-order matrix elements.
We thus write
\begin{widetext}
\begin{align}\label{soreg}
&\ 2\,\bigg\langle H^{(4)}_\mathrm{hfs}\,\frac{1}{(E_0-H_0)'}\,\bigg(\kappa\, H^{(4)}_\mathrm{nfs} + H^{(5)}_\mathrm{nfs}\bigg)\bigg\rangle
= E_{\sec,A} + E_{\mathrm{fo},A}\,.
\end{align}
After simple but tedious calculations we obtain
\begin{align}
E_{\sec,A} = &\
\frac{\alpha \,g}{9\,\pi\,M}\,\langle\vec I\cdot\vec S\rangle\,
\bigg\langle V_R\,\frac{1}{(E_0-H_0)'}\,\bigg[\alpha^2\bigg(\frac{5}{6}-\frac15+\ln\frac{\alpha^{-2}}{2\lambda}\bigg)V_R - \frac{7\,\alpha^2}{r^3} + 3\pi\,\kappa\,H_R\bigg]\bigg\rangle\
\end{align}
and
\begin{align}\label{foA}
E_{\mathrm{fo},A}
= &\ \langle\vec I\cdot\vec S\rangle\,\frac{g}{M}\bigg\{
 \frac{\alpha^3}{9\,\pi}\,\bigg[
 \bigg(\frac{5}{6}-\frac15+\ln\frac{\alpha^{-2}}{2\lambda}\bigg)\bigg(\big\langle 16\pi \,Z[\delta^3(r_1)+\delta^3(r_2)]\big\rangle\bigg\langle\frac{Z}{r_1}+\frac{Z}{r_2}\bigg\rangle
-\bigg\langle 16\pi\,Z[\delta^3(r_1)+\delta^3(r_2)]\nonumber\\
&\ \times\bigg(\frac{Z}{r_1}+\frac{Z}{r_2}\bigg)\bigg\rangle +2\bigg\langle\frac{Z^2}{r_1^4}+\frac{Z^2}{r_2^4}\bigg\rangle\bigg)
 - 14\bigg\langle\frac{1}{r^3}\bigg\rangle\bigg\langle\frac{Z}{r_1}+\frac{Z}{r_2}\bigg\rangle
+14\bigg\langle\frac{1}{r^3}\bigg(\frac{Z}{r_1}+\frac{Z}{r_2}\bigg)\bigg\rangle\bigg]\nonumber\\
&\ + \frac{\kappa\,\alpha}{3}\bigg[
\frac{\alpha}{4}\bigg\langle\frac{Z^2}{r_1^4}+\frac{Z^2}{r_2^4}
-2\bigg(\frac{Z\vec r_1}{r^3_1}-\frac{Z\vec r_2}{r^3_2}\bigg)\cdot\frac{\vec r}{r^3}\bigg\rangle
+\bigg\langle\bigg(\frac{Z}{r_1}+\frac{Z}{r_2}\bigg) (E_0-V)^2\bigg\rangle
-\frac{1}{2}\bigg\langle p_1^2\bigg(\frac{Z}{r_1}+\frac{Z}{r_2}\bigg)p_2^2\bigg\rangle\nonumber\\
&\ + 2\, E^{(4)}\bigg\langle\frac{Z}{r_1}+\frac{Z}{r_2}\bigg\rangle
+ \bigg\langle p_1^i\bigg(\frac{Z\alpha}{r_1}+\frac{Z\alpha}{r_2}\bigg)\bigg(\frac{\delta^{ij}}{r}+\frac{r^i r^j}{r^3}\bigg)p_2^j\bigg\rangle
 -\big\langle \pi\,Z\alpha\,\big[\delta^3(r_1)+\delta^3(r_2)\big]\big\rangle\bigg\langle\frac{Z}{r_1}+\frac{Z}{r_2}\bigg\rangle\bigg]\bigg\}\,,
\end{align}
\end{widetext}
where $E^{(4)}$ is the relativistic correction of order $m\alpha^4$ to the centroid energy.
The above formulas contain contributions from the self-energy and the vacuum polarization.
The latter is induced by the vacuum-polarization correction to the Coulomb potential,
\begin{eqnarray}
\delta V^{(1)} &=& -\frac{4\,\alpha^2}{15}Z\,\delta^3(r_1) + (1\leftrightarrow2)\,,
\end{eqnarray}
and is a part of the $H^{(5)}$ Hamiltonian in Eq.~(\ref{H5}).

\section{Matching the hydrogenic limit}
\label{app:hydrSE}

In this section we obtain the hydrogenic ($Z$$\to$$\infty$) limit of our formulas derived for the helium-like atom.
We first consider the normalized difference of hfs energies,
$\Delta E^{(7)} \equiv n^3\,E^{(7)}(nS) - E^{(7)}(1S)$,
for which the obtained limit should agree with the known results derived for the hydrogen-like atoms.
Next, we consider the $1S$ state, for which the limit of our formulas should differ
from the hydrogen result by a term proportional to the electron-nucleus
Dirac $\delta$ function. By matching these two results, we obtain the missing $\delta$-function
contribution.

We start with the self-energy part.
To get the hydrogenic limit of our formulas, we make the replacement $\vec S\rightarrow 2\vec s_1$ and
drop all the electron-electron terms containing $r$ and terms containing variables of the second electron.
For the normalized hfs difference,
omitting the low-energy part $E'_L$ and using known results for the expectation values of effective operators with hydrogenic wave functions,
we obtain
\begin{align}
&\ \Delta E^{(7)}({\rm se}) =  \frac{Z^6}{\pi}\frac{g}{2\,M}
\lbr\vec s_1\cdot\vec I\rbr \,
\bigg[\frac{71}{18} - \frac{79}{27n} - \frac{55}{54 n^2} \nonumber\\
&\ + \frac{214}{27} \big( \gamma + \Psi(n) - \ln(n)\big) + \big(-\ln(\alpha)^{-2}+\ln 2\big)\nonumber\\
&\ \times \bigg(-\frac{16}{3}
+ \frac{64}{9 n}
- \frac{16}{9 n^2} - \frac{64}{9} \big[\gamma + \Psi(n) - \ln(n)\big]\bigg)\bigg] \,.
\end{align}
This agrees with the result from Ref.~\cite{jentschura:06:hfs} after the replacement
$\ln(Z\alpha)^{-2}\rightarrow\ln(\alpha)^{-2}$,
which is caused by the different choices of the photon cutoff in the low-energy part.
For the $1S$ state, the hydrogenic limit of our formulas is
\begin{align}\label{ourground}
&\ E^{(7)}({\rm se},1S) =
\frac{Z^6}{\pi}\frac{g}{2\,M}\lbr\vec s_1\cdot\vec I\rbr \,
\bigg[-\frac{319}{18} + 16\ln\Lambda \nonumber\\
&\ + \frac{646}{27} \ln 2+ \frac{214}{27}\ln Z - \frac{64}{9} \big(\ln ^2 2 + \ln 2 \ln\Lambda   \nonumber\\
&\ + \ln Z\ln\Lambda  + \ln 2\ln Z\big)\bigg] + 2 Z^6\,\eta_\mathrm{se}\,\lbr\vec s_1\cdot\vec I\rbr \,,
\end{align}
where $\eta_\mathrm{se}$ parametrizes the missing $\delta$-function self-energy contribution.
The above result should agree with the hydrogenic $1S$ self-energy result
given by the sum of $F_M$ and $F_H$ in Eqs.~(A12) and (A13) of Ref.~\cite{patkos:23}.
In these equations the intermediate cutoff $\epsilon$ is the same as
cutoff $\Lambda = \alpha^2\lambda$ used in this work.
Matching Eq.~(\ref{ourground}) with the result from Ref.~\cite{patkos:23}, we get the missing
contribution
\begin{align}\label{deltamatched}
\eta_\mathrm{se} = &\ \frac{g}{4\,\pi\,M}
\bigg[-\frac76 - \frac{44\pi^2}{27} - \frac{10}{3}\zeta(3) + \frac{896}{27}\ln 2 \nonumber\\
&\ + \frac{16}{9}\ln^2 2
- \frac{64}{3}\ln\Lambda + \frac{16}{9}\ln^2\Lambda + \frac{160}{9}\ln 2\ln\Lambda \nonumber\\
&\ + \ln(\alpha)\bigg(\frac{214}{27} - \frac{64}{9}\ln\Lambda - \frac{64}{9}\ln 2\bigg)\bigg]\,.
\end{align}
Adding the corresponding contribution from the second electron,
then employing Eq.~(\ref{D9}) to rewrite their sum as a combination of the
symmetric and antisymmetric parts,
and dropping the antisymmetric part since it does not contribute for the $^3S$ state, we
get the complete $\delta$-function self-energy contribution for helium.

For the vacuum polarization we proceed in a similar fashion.
For normalized difference of $S$ states, the hydrogenic limit of our formulas,
\begin{align}
\Delta E^{(7)}({\rm vp}) = &\ -\frac{32\,Z^6\,g}{45\,\pi\,M}\,\lbr\vec s_1\cdot\vec I\rbr\,\bigg[\frac34 - \frac{1}{n}
\nonumber\\&\
+ \frac{1}{4n^2} +\gamma + \Psi(n) - \ln (n)\bigg]\,,
\end{align}
agrees with the result in Ref.~\cite{jentschura:06:hfs}.
For the $1S$ state, the hydrogenic limit of our formulas,
\begin{align}
E^{(7)}({\rm vp},1S) = &\ \frac{Z^6}{\pi}\frac{2g}{45\,M}\lbr\vec s_1\cdot\vec I\rbr \,
\big(36-16\ln2-16\ln Z\big) \nonumber\\
&\ + 2 Z^6\,\eta_\mathrm{vp}\,\lbr\vec s_1\cdot\vec I\rbr \,,
\end{align}
is matched with the literature result \cite{eides:01},
yielding the missing $\delta$-function contribution
\begin{equation}
\eta_\mathrm{vp} = -\frac{g}{45\pi\,M}\,\bigg[\frac{472}{15} + 16\ln(\alpha)\bigg]\,.
\end{equation}
It should be noted that for the vacuum polarization
we checked this result by a direct derivation of the $\delta$-function contribution
by using the dimensional regularization.

The sum of the self-energy and vacuum-polarization $\delta$-function contributions yields the
term $E_{{\rm fo}, B}$ given by Eqs.~(\ref{eq:eta}) and (\ref{eq:eta2}).

\section{Cancellation of $\lambda$-dependent terms}
\label{app:lambda}

In this section we obtain the final formulas for the $m\alpha^7$ correction
and demonstrate the cancellation of terms depending on the low-energy cutoff $\lambda$.
We start with the self-energy contribution, which is represented as a sum of three terms,
\begin{align}
E^{(7)}(\rm se) = &\ E_L + E_{\rm sec}({\rm se}) + E_{\rm fo}({\rm se})\,.
\end{align}
Here,
$E_L$ is the Bethe-logarithm low-energy contribution given by Eq.~(\ref{ELl}).
The second-order self-energy contribution $E_{\rm sec}(\rm se)$ is obtained in Appendix~\ref{app:2ndorder} as
\begin{widetext}
\begin{align}\label{sofin}
E_{\rm sec}({\rm se}) = &\ \frac{g}{2\,\pi\,M}\bigg\{
\frac{2 }{9}\,\langle\vec I\cdot\vec S\rangle\,
\bigg\langle V_R\,\frac{1}{(E_0-H_0)'}\,\bigg[\bigg(\frac{5}{6}+\ln\frac{\alpha^{-2}}{2\lambda}\bigg)V_R - \frac{7}{r^3} + \frac32 H_R\bigg]\bigg\rangle
\nonumber\\
&\ + 2Z\,\bigg\langle \bigg[\frac{\vec r_1}{r_1^3}\times\vec p_1+\frac{\vec r_2}{r_2^3}\times\vec p_2\bigg]\cdot\vec I\,
\frac{1}{(E_0-H_0)'}\,\bigg[
\frac{Z}{4}\biggl(\frac{\vec{ r}_1}{r_1^3}\times\vec{ p}_1+\frac{\vec{ r}_2}{r_2^3}\times\vec{ p}_2\biggr)
-\frac{1}{2}\,\frac{\vec{ r}}{r^3}\times(\vec{ p}_1-\vec{ p}_2)\bigg]\cdot\vec S\bigg\rangle\nonumber\\
&\ -\frac{3}{2}Z\bigg\langle  S^i I^j\bigg[\frac{1}{r_1^3}\bigg(\delta^{ij} - 3 \frac{r_1^i r_1^j}{r_1^2}\bigg)
+\frac{1}{r_2^3}\bigg(\delta^{ij} - 3 \frac{r_2^i r_2^j}{r_2^2}\bigg)\bigg]\,\frac{1}{(E_0-H_0)'}\,\frac{1}{4}\left(
\frac{\vec{\sigma}_1\,\vec{\sigma}_2}{r^3}
-3\,\frac{\vec{\sigma}_1\cdot\vec{r}\,\vec{\sigma}_2\cdot\vec{r}}{r^5}\right)\bigg\rangle\bigg\}\,.
\end{align}
The first-order self-energy contribution reads
\begin{align}
E_{\rm fo}({\rm se}) \label{E3}
  = &\ \frac{g}{4\,\pi\,M}\,\langle\vec I\cdot\vec S\rangle\,\bigg[\frac{2}{9}\,
 \bigg(\frac{71}{3}+32\ln\frac{\alpha^{-2}}{2\lambda}\bigg)\big\langle \pi \,Z\,\delta^3(r_1)\big\rangle\bigg\langle\frac{Z}{r_1}+\frac{Z}{r_2}\bigg\rangle
 +\bigg(\frac{143}{108} + \frac89\ln\frac{\alpha^{-2}}{2\lambda}\bigg)\bigg\langle \frac{Z^2}{r_1^4}\bigg\rangle\nonumber\\
&\ -\frac{2 }{3}\,
 \bigg(\frac{85}{6}+16\ln\frac{\alpha^{-2}}{2\lambda} \bigg)\bigg\langle \pi\,Z^2\,\delta^3(r_1)\frac{1}{r_2}\bigg\rangle
 - \frac{56}{9}\bigg\langle\frac{1}{r^3}\bigg\rangle\bigg\langle\frac{Z}{r_1}\bigg\rangle\nonumber\\
&\
+\frac{56}{9}\bigg\langle\frac{Z}{r^3\,r_1}\bigg\rangle
-\frac{13}{12}\bigg\langle \frac{Z\,\vec r_1}{r^3_1}\cdot\frac{\vec r}{r^3}\bigg\rangle
 + \frac{4Z}{3}\, E^{(4)}\bigg\langle\frac{1}{r_1}\bigg\rangle
 +\frac{2Z}{3}\bigg\langle\frac{1}{r_1} (E_0-V)^2\bigg\rangle\nonumber\\
&\
-\frac{Z}{3}\bigg\langle p_1^2 \frac{1}{r_1} p_2^2\bigg\rangle
 + \frac{2Z}{3}\bigg\langle p_1^i\,\frac{1}{r_1}\bigg(\frac{\delta^{ij}}{r}+\frac{r^i r^j}{r^3}\bigg)p_2^j\bigg\rangle
+\frac{Z}{9}\bigg(\frac{77}{6} + 16 \ln\frac{\alpha^{-2}}{2\lambda} \bigg)
 \langle p_1^k \,\pi\,\delta^3(r_1)\,p_1^k\rangle\nonumber\\
&\ -\frac{\pi\,Z}{9}\bigg(\frac{95}{3} + 32 \ln\frac{\alpha^{-2}}{2\lambda} \bigg)
\bigg\langle\bigg(E_0-\frac{1}{r_2}-\frac{p_2^2}{2}\bigg) \,\delta^3(r_1)\bigg\rangle
\bigg]
 + \langle\vec I\cdot\vec S\rangle\,\eta_\mathrm{se}\,Z^3\,\pi\,\langle\delta^3(r_1)\rangle + (1\leftrightarrow2)\,.
\end{align}
\end{widetext}
For the simplification of the result we used the identity
\begin{align}
&\ \bigg\langle p_1^i\,\frac{1}{r_1^3}\bigg(\delta^{ij}-3\frac{r_1^i r_1^j}{r_1^2}\bigg)\,p_1^j\bigg\rangle
=  \bigg\langle\frac23\, p_1^k \,Z\,\pi\,\delta^3(r_1)\,p_1^k + \frac{Z^2}{r_1^4} \nonumber\\
&\ - \frac{Z\,\vec r_1}{r^3_1}\cdot\frac{\vec r}{r^3}
-\bigg(E_0+\frac{Z-1}{r_2}-\frac{p_2^2}{2}\bigg) Z\,\pi\delta^3(r_1)\bigg\rangle\,.
\end{align}

By algebraic calculations we checked that the dependence on the photon  momentum cutoff $\lambda$ is
canceled in the sum of  $E_L$, $E_{\rm sec}(\rm se)$, and $E_{\rm fo}(\rm se)$. After that, we can remove the
$\lambda$ dependence by setting $\lambda=1$ in all formulas.
In this way we obtain the final formulas
given by Eqs.~(\ref{fofinQ}) and (\ref{sofin2}).
The second-order contribution in Eq.~(\ref{sofin2}) is obtained from Eq.~(\ref{sofin})
after spin averaging with the help of a formula,
\begin{align}\label{spinav2}
\langle S^i I^j Q_1^{ij} \, \sigma_1^a \sigma_2^b\,Q_2^{ab}\rangle =&\,
\frac{\lbr\vec I\cdot\vec S\rbr}{3}\,\lbr Q_1^{ij} Q_2^{ij} \rbr\,,
\end{align}
in the second and the third line of Eq.~(\ref{sofin}), correspondingly.

The final result for the vacuum-polarization contribution
is a sum of the corresponding parts of the second-order contribution in Eq.~(\ref{soreg}),
the first-order contribution contained in $H^{(7)}_\mathrm{hfs}$ in Eq.~(\ref{H7se}),
and the additional Dirac-$\delta$-like part obtained in Appendix~\ref{app:hydrSE}. We thus get
\begin{align}\label{finvp}
&\ E^{(7)}({\rm vp}) = E_{\rm sec}({\rm vp}) + E_{\rm fo}(\rm{vp}) \,,\\
&\ E_{\rm sec}({\rm vp}) = -\frac{g}{45\,\pi\,M}\,\lbr\vec I\cdot\vec S\rbr\,
\bigg\langle V_R\,\frac{1}{(E_0-H_0)'}\,V_R \bigg\rangle\,,\label{E7}\\
&\ E_{\rm fo}({\rm vp}) = -\frac{g}{45\,\pi\,M}\,\lbr\vec I\cdot\vec S\rbr\,
  \bigg\{\big\langle 16\pi \,Z \, \delta^3(r_1)\big\rangle\bigg\langle\frac{Z}{r_1}+\frac{Z}{r_2}\bigg\rangle \nonumber \\
&\ +4\,\big\langle\vec p_1\,\pi\,Z\,\delta^3(r_1)\,\vec p_1\big\rangle
+\bigg\langle 8\pi\,(Z-3Z^2)\, \delta^3(r_1)\frac{1}{r_2}\bigg\rangle\nonumber\\ &\
- 8\,\bigg\langle\bigg(E_0-\frac{p_2^2}{2}\bigg)\pi\,Z\,\delta^3(r_1)\bigg\rangle
+2\bigg\langle\frac{Z^2}{r_1^4}\bigg\rangle\nonumber\\
&\
+ \bigg(\frac{472}{15} + 16\ln(\alpha)\bigg)\,\big\langle \pi\,Z^3\,\delta^3(r_1)\big\rangle
+ (1\leftrightarrow2)\bigg\}\,. \label{E8}
\end{align}

\end{document}